# Chemical Composition Tuning of the Anomalous Hall Effect in Isoelectronic L1$_0$ FePd$_{1-x}$Pt$_x$ Alloy Films

P. He,[1,2] L. Ma,[1] Z. Shi,[1] G. Y. Guo,[3,4,*] J.-G. Zheng,[5] Y. Xin,[6] and S. M. Zhou[1,2,†]

[1] *Department of Physics, Tongji University, Shanghai 200092, P. R. China*
[2] *Department of Physics, Fudan University, Shanghai 200433, P. R. China*
[3] *Graduate Institute of Applied Physics, National Chengchi University, Taipei 11605, Taiwan*
[4] *Department of Physics, National Taiwan University, Taipei 10617, Taiwan*
[5] *Laboratory for Electron and X-ray Instrumentation, University of California, Irvine, CA 92697-2800, USA*
[6] *NHMFL, Florida State University, Tallahassee, FL 32310, USA*
(Dated: May 19, 2012)

The anomalous Hall effect (AHE) in L1$_0$ FePd$_{1-x}$Pt$_x$ alloy films is studied both experimentally and theoretically. We find that the intrinsic contribution ($\sigma_{AH}^{int}$) to the AHE can be significantly increased whereas the extrinsic side-jump contribution ($\sigma_{AH}^{sj}$) can be continuously reduced from being slightly larger than $\sigma_{AH}^{int}$ in L1$_0$ FePd to being much smaller than $\sigma_{AH}^{int}$ in L1$_0$ FePt, by increasing the Pt composition $x$. We show that this chemical composition tuning of the intrinsic contribution is afforded by the stronger spin-orbit coupling strength on the Pd/Pt site when the lighter Pd atoms are replaced by the heavier Pt atoms. Our results provide a means of manipulating the competing AHE mechanisms in ferromagnetic alloys for fuller understanding the AHE and also for technological applications of ferromagnetic alloys.

Anomalous Hall effect (AHE) refers to the transverse charge current generation in solids in a ferromagnetic phase generated by the electric field. The AHE, though first discovered by Hall in 1881[1], has received intensive renewed interest in recent years mainly because of its close connection with spin transport phenomena[2]. There are several competing mechanisms proposed for the AHE. Extrinsic mechanisms of skew scattering[3] and side jump[4] result from the asymmetric impurity scattering caused by the spin-orbit interaction (SOI). Another mechanism arises from the transverse velocity of the Bloch electrons induced by the SOI, discovered by Karplus and Luttinger[5]. This intrinsic AHE has recently been reinterpreted in terms of the Berry curvature of the occupied Bloch states.[6] Experimentally, the measured anomalous Hall resistivity $\rho_{AH}$ is often analyzed in terms of two distinctly different resistivity ($\rho_{xx}$)-dependent terms[2], i.e.,

$$\rho_{AH} = a\rho_{xx} + b\rho_{xx}^2. \qquad (1)$$

Since usually $\rho_{AH} \ll \rho_{xx}$, the anomalous Hall conductivity (AHC) $\sigma_{AH} = \rho_{AH}/(\rho_{AH}^2 + \rho_{xx}^2) \approx a\sigma_{xx} + b$, where the linear $\sigma_{xx}$-dependent term ($a\sigma_{xx}$) was attributed to the extrinsic skew scattering mechanism ($\sigma_{AH}^{sk}$)[3]. The skew scattering contribution has been found to become dominant in dilute impurity metals at low temperatures.[2] The scattering-independent term $b$ was further separated into the intrinsic contribution ($\sigma_{AH}^{int}$)[5] which can be obtained from band structure calculations[2], and the extrinsic side jump mechanism ($\sigma_{AH}^{sj}$)[4], i.e., $b = \sigma_{AH}^{int} + \sigma_{AH}^{sj}$.

Several fundamental issues concerning the AHE remain unresolved, despite intensive theoretical and experimental studies in recent years. First-principles studies based on the Berry phase formalism showed that the intrinsic AHC agrees well with the measured scattering-independent contribution $b$ in various materials.[7,8] This indicates that the side jump contribution is small, and hence the intrinsic mechanism can be assumed to be dominant[2]. This is supported by the recent experimental finding of the negligible $\sigma_{AH}^{sj}$ in fcc Ni[9] using a newly proposed empirical $\sigma_{xx}$-scaling formula for $\sigma_{AH}$[10]. However, recent theoretical calculations for the 2D Rashba and 3D Luttinger Hamiltonians using a Gaussian disorder model potential suggested that the AHE in the (III,Mn)V ferromagnetic semiconductors at low temperatures could be dominated by the $\sigma_{AH}^{sj}$.[11] Perhaps, most surprising is the recent report of the side jump dominant mechanism in L1$_0$ FePd but the intrinsic dominant mechanism in L1$_0$ FePt.[12] This finding was based on the $\sigma_{AH}^{sj}$ deduced as the difference between the measured $b$ and calculated $\sigma_{AH}^{int}$, and thus hinges on the accurate determination of both experimental $b$ and theoretical $\sigma_{AH}^{int}$. However, the latest *ab initio* calculations[13] gave the $\sigma_{AH}^{sj}$ that is about three times smaller than the deduced $\sigma_{AH}^{sj}$ in Ref. 12. This confusing situation on the side jump contribution to the AHC is partially caused by the fact that the $\sigma_{AH}^{sj}$ is independent of the impurity concentration although it arises from the SOI-induced asymmetric electron scattering at the impurities.[2,4,14]

All the above mentioned AHE mechanisms are caused by relativistic SOI in solids.[2] In particular, *ab initio* calculations showed that the intrinsic AHC depends almost linearly on the SOI strength ($\xi$) in, e.g., bcc Fe[7], L1$_0$ FePd and L1$_0$ FePt[12]. For better understanding of the AHE and related spin transport phenomena, it is very important to be able to experimentally tune the relative contributions of the different mechanisms to the AHC via systematic SOI engineering. To this end, isoelectronic L1$_0$ FePd$_{1-x}$Pt$_x$ (FePdPt) ternary alloy films are ideal candidates because $\xi$ could be smoothly increased several times by gradually replacing the Pd atoms with the Pt atoms while keeping other physical parameters such as crystalline structure and lattice constants almost unchanged. Therefore, extensive studies of the AHE in L1$_0$ FePdPt ternary alloys with $x$ ranging from 0.0 to 1.0 will be crucial to establish an overall physical picture on the $\xi$ dependence of various mechanisms and also to clarify the role of the extrinsic side jump scattering. Furthermore, these ternary alloys have promising applications to both high performance permanent magnets and ultrahigh density magnetic recording media[15] because of their high uniaxial magnetic anisotropy.

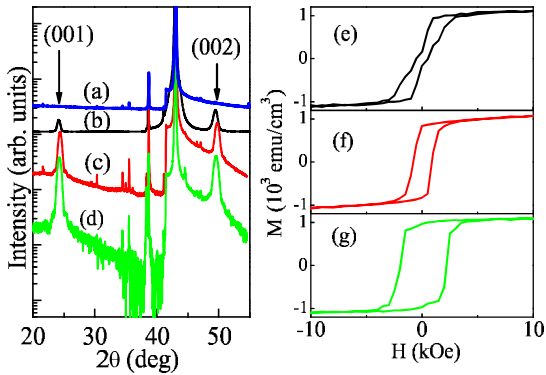

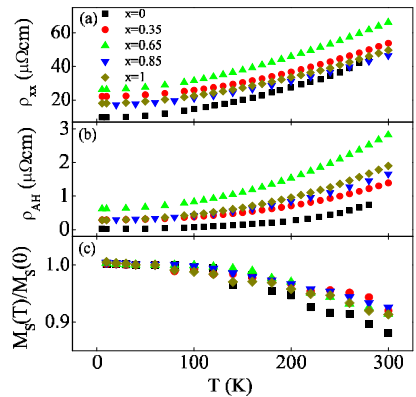

FIG. 1: (color online) XRD patterns (b, c, d) and magnetization hysteresis loops (e, f, g) for L1$_0$ FePd$_{1-x}$Pt$_x$ alloys with $x = 0$ (b, e), 0.65 (c, f), and 1.0 (d, g). In (a), XRD pattern of MgO(002) substrate is given.

FIG. 2: (color online) $\rho_{xx}$ (a), $\rho_{AH}$ (b), and normalized spontaneous magnetization (c) for L1$_0$ FePd$_{1-x}$Pt$_x$ alloys.

In this Letter, we report the different contributions to the AHC in isoelectronic L1$_0$ FePtPd alloy films. Here, we succeeded in fabricating highly ordered L1$_0$ FePdPt ternary alloy films for several different composition ratios of Pd to Pt atoms. By carefully designed experimental procedure, we determined the scattering-independent component $b$ and skew scattering contribution for these films using the measured longitudinal and Hall resistivities as well as magnetization over a wide temperature range via the scaling formula. We performed relativistic band structure calculations to determine the intrinsic contribution. We obtained the side jump contribution by subtracting the theoretical intrinsic contribution from the experimental scattering-independent component. Our results help to better understand the competing mechanisms of the AHE in ferromagnetic alloys, a controversial issue of considerable current interest in solid state physics.

L1$_0$ FePd$_{1-x}$Pt$_x$ films with several different $x$ values were deposited by DC magnetron sputtering. Details of sample fabrication, characterization of film thickness and microstructure, and measurements of magnetic properties and the anomalous Hall resistivity $\rho_{AH}$ and longitudinal resistivity $\rho_{xx}$ are described elsewhere[16]. The film thickness is $20 \pm 1$ nm. The x-ray diffraction (XRD) peak around $2\theta = 23°$ indicates the establishment of the long range chemical ordering, as shown in Figs. 1(b)-1(d). Since XRD peaks in the $2\theta$ region from $30°$ to $40°$ come from MgO substrate, all samples are of L1$_0$ single phase. The epitaxial quality of all samples is confirmed by high resolution transmission electronic microscopy results[16]. For L1$_0$ FePd and FePt films, the ordering parameter $S = 0.81$ and 0.71, respectively. For $0 < x < 1$, $S$ lies between 0.7 and 0.8. As shown in Figs. 1(e)- 1(g), the perpendicular magnetic anisotropy in the L1$_0$-ordered FePdPt films could be tuned by varying $x$[17,18].

The measured $\rho_{xx}$, $\rho_{AH}$ and spontaneous magnetization $M_S$ are displayed as a function of temperature ($T$) in Fig. 2. We find that the $\rho_{xx}(T)$ can be fitted by a linear function of $T^2$, indicating the dominant electron scattering by spin flip[19,20]. Also, the $\rho_{AH}$ is about two orders of magnitude smaller than $\rho_{xx}$. The residual resistivity $\rho_{xx0}$ [i.e., $\rho_{xx}(T=0)$] changes non-monotonically with $x$ and reaches the maximal value at $x = 0.65$. This may be attributed to the different crystalline quality for different Pd contents[17,18]. Figure 2(c) shows that $M_S$ decreases with increasing $T$. For all samples, the $M_s(T)$ can be fitted by a linear function of $T^2$, being due to either the excitation of long wavelength spin waves or the interaction between spin waves or both[8,21].

To extract different contributions to the AHC using the scaling formula of Eq. (1), the $\rho_{AH}/\rho_{xx}$ ratio is plotted as a function of $\rho_{xx}$ in Fig. 3(a). Clearly, the $\rho_{AH}/\rho_{xx}$ ratio deviates from the linear $\rho_{xx}$-dependence. This deviation arises from our experimental strategy that for a given sample, the variations of $\rho_{xx}$ and $\rho_{AH}$ are accomplished by varying temperature. However, varying temperature inevitably changes $M_S$ in the sample. In Fe films which have a Curie temperature ($T_C$) above 1000 K, $M_S$ hardly changes with $T$ below room temperature[10]. In contrast, $L1_0$ FePdPt films have a lower $T_C$ of around 740 K[22], and thus the measured $M_S$ varies significantly with $T$, as shown in Fig. 2(c). Previous *ab initio* calculations and experimental studies[8,23,24] showed that because the intrinsic AHC is proportional to $M_S$, the $\sigma_{AH}^{int}$ and $M_S$ have the same $T$-dependence [$f(T)$]. Also, the skew scattering contribution $\sigma_{AH}^{sk}$ was found to be roughly proportional to $M_S$[14]. Therefore, when using the scaling formula of Eq. (1) to fit the measured AHC, we assume $a = a_0 f(T)$ and $b = b_0 f(T)$ where $a_0$ and $b_0$ refer, respectively, to $a$ and $b$ at 0 K. The validity of the above assumption is evident in Fig. 3(b) where the $\rho_{AH}/[\rho_{xx}f(T)]$ ratio for all the samples is shown to be a perfect linear function of $\rho_{xx}$. The slope and the intercept of the lines are the $b_0$ and $a_0$, respectively. This shows that, to reliably deduce $b_0$ and $a_0$ for $L1_0$ FePdPt alloy films, it is essential to take into account the $T$-dependence of the $M_S$ in the $T$ regime studied here.

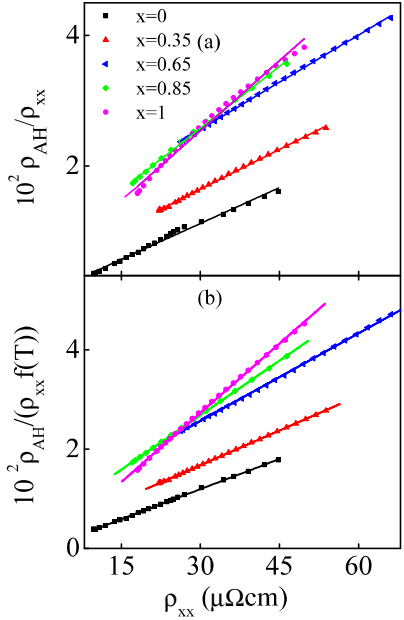
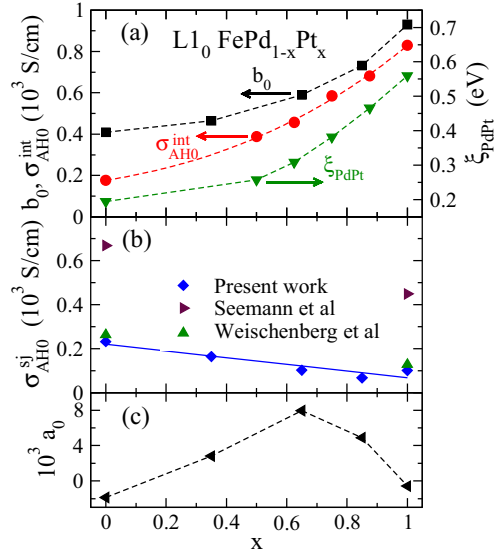

FIG. 3: (color online) $\rho_{AH}/\rho_{xx}$ (a) and $\rho_{AH}/[\rho_{xx}f(T)]$ (b) versus $\rho_{xx}$ for $L1_0$ FePd$_{1-x}$Pt$_x$ films. Solid lines in (b) refer to a linear fit.

FIG. 4: (color online) Experimental $b_0$, theoretical $\sigma_{AH0}^{int}$ and $\xi_{PdPt}$ (a), deduced $\sigma_{AH0}^{sj}$ (b), and experimental $a_0$ (c) for $L1_0$ FePd$_{1-x}$Pt$_x$ films as a function of the Pt concentration $x$. In (b), the $\sigma_{AH0}^{sj}$ values estimated by Seemann et al.[12] and calculated from first-principles by Weischenberg et al.[13], are also displayed. The solid line in (b) is a linear fit.

To determine the intrinsic AHC, we perform self-consistent relativistic band structure calculations within the density functional theory with the generalized gradient approximation[25] for pure $L1_0$ FePd and FePt as well as $L1_0$ FePd$_{1-x}$Pt$_x$ alloys in the virtual crystal approximation[16]. The calculated $\sigma_{AH0}^{int}$ [displayed in Fig. 4(a)] for $L1_0$ FePd and FePt is 177 and 830 S/cm, respectively, being in rather good agreement with previous *ab initio* calculations[12]. Interestingly, Fig. 4(a) shows that both the experimentally derived scattering-independent term $b_0$ and theoretical $\sigma_{AH0}^{int}$ increase monotonically with Pt concentration $x$. The $b_0$ ($\sigma_{AH0}^{int}$) is 409 (177) and 930 (830) S/cm, respectively, for $L1_0$ FePd ($x = 0$) and $L1_0$ FePt ($x = 1$) films. This demonstrates that the AHC in the FePdPt ternary alloys, can be engineered by chemical composition tuning. To understand the mechanism of this chemical composition tuning, we display in Fig. 4(a) the calculated Pd/Pt-site $d$-orbital SOI strength $\xi_{PdPt}$ as a function of $x$. The calculated Fe $d$-orbital SOI strength $\xi_{Fe}$ of 0.061 eV is independent of $x$. The calculated $\xi_{Pt}$ ($x = 1$) is 0.559 eV, being almost three times larger than the calculated $\xi_{Pd}$ ($x = 0$) of 0.194 eV. Fig. 4(a) shows that the variations of both $b_0$ and $\sigma_{AH0}^{int}$ with $x$ follow closely that of $\xi_{PdPt}(x)$. Clearly, the chemical composition tuning of the AHC in $L1_0$ FePdPt films reported here is realized by substituting the Pd atoms with the heavier Pt atoms which have a stronger SOI strength.

As mentioned above, $b_0 = \sigma_{AH0}^{int} + \sigma_{AH0}^{sj}$ where $\sigma_{AH0}^{int}$ and $\sigma_{AH0}^{sj}$ denote $\sigma_{AH}^{int}$ and $\sigma_{AH}^{sj}$ at 0 K, respectively. Thus, using the measured $b_0$ and calculated $\sigma_{AH0}^{int}$ here, we can estimate $\sigma_{AH0}^{sj}$ for the $L1_0$ FePd$_{1-x}$Pt$_x$ alloys. To determine the $\sigma_{AH0}^{int}$ for the experimental Pt concentrations, we fit a polynomial to the calculated $\sigma_{AH0}^{int}$ values and find that $\sigma_{AH0}^{int} = 177 + 230x + 309x^2 + 113x^3$ S/cm gives a good fit, as the red dashed curve in Fig. 4(a) shows. The $\sigma_{AH0}^{sj}$ estimated this way is displayed in Fig. 4(b). We find $\sigma_{AH0}^{sj}$ for the pure $L1_0$ FePd and $L1_0$ FePt films to be 232 and 101 S/cm, respectively. These

$\sigma_{AH0}^{sj}$ values are in good agreement with the corresponding theoretical values of 263 and 128 S/cm from latest *ab initio* calculations[13]. Interestingly, Fig. 4(b) shows that $\sigma_{AH0}^{sj}$ decreases slightly with increasing $x$. Above the composition $x_c = 0.11$, $\sigma_{AH0}^{sj}$ is smaller than $\sigma_{AH0}^{int}$.

Figure 4(c) shows that $a_0$ changes non-monotonically with $x$ and reaches a maximal value of 0.008 at $x = 0.65$. This may be attributed to the non-monotonic variation of defect concentration (grain boundaries, line defects, microtwins, point defects and so on) with $x$, as is demonstrated by the variation of $\rho_{xx0}$ in Fig. 2(a). It may also arise from the artifact that both the PtPd site disorder in the L1$_0$ structure and the scattering reach the maxima at intermediate $x$. Similar correlation between the $a_0$ and the impurity concentration was also observed for Fe films with different thickness[10] and L1$_0$ FePt films treated with different annealing conditions[27]. In the latter case, during the establishment of the chemical ordering at elevated annealing temperatures, the impurity concentration is reduced due to the improved crystalline quality, leading to the reduced residual resistivity and $a_0$.

Different $\xi$ dependences of the $\sigma_{AH0}^{int}$ and $\sigma_{AH0}^{sj}$ may reflect a strong difference in their distribution on the Fermi surface[13]. The $\sigma_{AH0}^{int}$ is simply given as the Brillouin zone integral of the Berry curvature over the pristine crystal's occupied electronic states[7] and is mainly contributed by certain "hot loops" in the vicinity of the intersections between different sheets of bands[13]. When the spin-orbit splitting occurs near the Fermi surface, very sharp peaks of the Berry curvature and the large $\sigma_{AH0}^{int}$ are induced. For L1$_0$-ordered alloy films with the magnetization along the $c$-axis, the two degenerate $t_{2g}$ orbitals $d_{yz}$ and $d_{zx}$ are lifted when $\xi$ on the Pd/Pt atomic site is enhanced with increasing $x$, leading to a significant increase of $\sigma_{AH0}^{int}$[28]. In contrast, since the $\sigma_{AH0}^{sj}$ does not contain singularities near those band crossings but is mainly contributed by certain isolated "hot spots", it does not show a remarked sensitivity on $\xi$. Finally, the variation trends of the $\sigma_{AH0}^{sj}$ and $\sigma_{AH0}^{int}$ with $x$ are again in agreement with the results based on massive Dirac Hamiltonian with randomly distributed weak $\delta$-function-like spin-independent impurities[14], where the $\sigma_{AH0}^{int}/\sigma_{AH0}^{sj}$ ratio is expected to change as a linear function of $\xi^2$.

In summary, we have determined the competing contributions to the AHC in isoelectronic L1$_0$ FePd$_{1-x}$Pt$_x$ alloy films. We found that the scattering independent component (intrinsic contribution) to the AHC can be continuously tuned from 409 (177) S/cm for L1$_0$ FePd ($x = 0$) to 930 (830) S/cm for L1$_0$ FePt ($x = 1$) by increasing the Pt composition $x$ whereas the extrinsic side-jump contribution decreases slightly from being slightly larger than $\sigma_{AH}^{int}$ at $x = 0$ to being much smaller than $\sigma_{AH}^{int}$ at $x = 1$. We have related this chemical composition tuning of the intrinsic contribution to the modification of the SOI strength on the Pd/Pt site when the lighter Pd atoms are replaced by the heavier Pt atoms. Our results would help better understand the origins of various mechanisms of the AHE in ferromagnetic alloys and also pay the way for designing the broad class of ferromagnetic L1$_0$ XPdPt (X=Fe and Co) ternary alloys[29,30] for spintronic devices and magnetic sensors.

This work was supported by the National Natural Science Foundation of China, the National Basic Research Program of China, the National Science Council of Taiwan, FSU Research Foundation and NHMFL (NSF DMR-0654118).

* Electronic address: gyguo@phys.ntu.edu.tw
† Electronic address: shiming@tongji.edu.cn